\documentclass[twocolumn,showpacs,preprintnumbers,amsmath,amssymb]{revtex4}

\usepackage{graphicx}
\usepackage{dcolumn}
\usepackage{bm}

\begin{document}


\title{Non-adiabatic dynamics in $^{10}$Be 
with the microscopic $\alpha$+$\alpha$+$N$+$N$ model
}

\author{M. Ito
}
\affiliation{%
Institute of Physics, University of Tsukuba, 305-8571 Tsukuba, Japan
}%

\date{\today}

\begin{abstract}
The $\alpha$+$^6$He low-energy reactions and the 
structural changes of $^{10}$Be in the microscopic 
$\alpha$+$\alpha$+$N$+$N$ model 
are studied by the generalized two-center cluster model with 
the Kohn-Hulth\'en-Kato variation method. It is found that, 
in the inelastic scattering to the $\alpha$+$^6$He(2$_1^+$) channel, 
characteristic enhancements are expected as the results of the 
parity-dependent non-adiabatic dynamics. In the positive parity state, 
the enhancement originates from the excited eigenstate generated by 
the radial excitation of the relative motion between two $\alpha$-cores. 
On the other hand, the enhancement in the negative parity state 
is induced by the Landau-Zener level-crossing. These non-adiabatic 
processes are discussed in connection to the formation of the inversion 
doublet in the compound system of $^{10}$Be. 
\end{abstract}

\maketitle
In the last two decades, developments of experiments with secondary 
RI beam have extensively proceeded the studies on light neutron-rich 
nuclei. In particular, much efforts have been devoted to the investigation 
of molecular structure in Be isotopes. Theoretically, molecular models
with $\pi$ and $\sigma$ orbitals along the axis connecting two 
$\alpha$-particles have been successful in understanding the low-lying 
states of this isotopes \cite{ITA,ENYO}. 
Experimentally, the molecular structures
were mainly investigated by the breakup processes \cite{SAITO,AHME} 
and the sequential decays \cite{MILI} using the high energy RI beams. 

In recent experiments, furthermore, the low-energy $^6$He beam becomes 
available. Low-energy reaction cross-sections such as 
the elastic scattering with an $\alpha$ target \cite{RAAB,SHIM} 
and the sub-barrier fusions with heavy target \cite{FUSE} have been 
accumulated. In future experiments, it will also be possible to 
investigate the molecular states in Be isotopes through the reactions 
such as $\alpha$+$^6$He and $\alpha$+$^8$He with low-energy 
$^{6,8}$He-beams. Therefore, it is very interesting to study theoretically 
on the low-energy scattering of $^6$He and $^8$He by an 
$\alpha$ target. 

In studying reaction processes exciting the molecular degrees of 
freedom, it is very important to construct a unified model which is 
capable of describing both structure and reaction on the same footing. 
For this purpose, we introduce a microscopic model, the generalized 
two-center cluster model (GTCM) \cite{GTCM1,GTCM2}. In this model, it is 
possible to describe both molecular and atomic limit of the system of 
$C_1$+$C_2$+$N$+$N$+... where $C_i$ is the $i$-th cluster core and $N$ 
is the nucleon. In the region where two core nuclei are close, 
the total system is expected to form the molecular orbital structure, 
while in the region where two core nuclei are far apart, the molecular 
orbitals smoothly change into product wave functions consisting of 
the atomic orbitals. 

In this paper, we apply the GTCM for the $^{10}$Be nucleus with 
the $\alpha$+$\alpha$+$N$+$N$ four-body model. 
We will analyze both 
molecular structure in $^{10}$Be and the low-energy $\alpha$+$^6$He 
scattering. Besides the description of the $\alpha$+$^6$He 
reaction, such analysis will be useful 
to understand the breakup mechanism of $^{10}$Be into clusters. 
In spite of many theoretical efforts in the last decade
\cite{ITA,ENYO},\cite{ARAI}--\cite{4body2}, 
only Ref.~\cite{ARAI} discusses the molecular-orbital 
formation in $^{10}$Be and the $\alpha$+$^6$He scattering 
problem in a unified way. 

Current experimental investigations are extended 
to the $^{12}$Be and $^{14}$Be nuclei \cite{SAITO,SHIM} and hence, 
theoretical studies extended to such heavier systems are urged. 
The GTCM approach has a potential to $^{12}$Be and $^{14}$Be as 
well as $^{10}$Be. On the other hand, the direct extension of the 
approach in Ref.~\cite{ARAI} becomes quite difficult for systems 
with many valence neutrons, since it utilizes the 
$\alpha$+$\alpha$+$N$+$N$ few-body model. 

One of the favorable features of our approach is that 
it is possible to describe the nuclear Landau-Zener (L-Z) 
transition microscopically \cite{ABE}-\cite{IMAN}. 
The L-Z transition is 
induced by the avoided level-crossing of two adiabatic 
potential-surfaces. When the avoided crossing occurs in 
the potential surfaces, the adiabatic states drastically change 
their intrinsic character at the crossing point. 
For large relative velocity, the colliding nuclei 
follows not the adiabatic path but the diabatic one. 
This diabatic motion was discussed by Landau and Zener for 
the atomic collision \cite{LANDAU}. The nuclear analogue has 
been called the nuclear L-Z transition. In spite of long 
history on this issue, a clear evidence of the L-Z transition 
is still lacking in nuclear collision \cite{EXP1,EXP2}. 
Our microscopic approach will indicate a possible nuclear L-Z 
transition in the collision of $\alpha$+$^6$He.

First, we briefly explain the framework of GTCM \cite{GTCM1,GTCM2}. 
The basis functions for $^{10}$Be are given as 
\begin{eqnarray}
\Phi^{J^\pi K}_{m,n}(S)
\;&=&\;\hat{P}^{J^\pi}_K\cdot{\cal A}\left\{\psi_L(\alpha)
\psi_R(\alpha)\varphi(m) 
\varphi(n)\right\}\;\;. 
\label{aobasis}
\end{eqnarray}
The $\alpha$-cluster wave function 
$\psi_i(\alpha)$ ($i$=$L, R$) is given by the (0s)$^4$ configuration 
in the harmonic oscillator (HO) potential. The position of an 
$\alpha$-cluster is explicitly specified as the 
left (L) or right (R) side. The relative motion between $\alpha$ particles 
is described by a localized Gaussian function specified by the distance 
$S$ \cite{HORI}. 
A single-particle state for valence neutrons around one of $\alpha$ 
clusters is given by an atomic orbitals (AO), 
$\varphi(i,p_k,\tau)$ with the subscripts of a center 
$i$ (=$L$ or $R$), a direction $p_k$ ($k$=$x$, $y$, $z$) of 
0$p$-orbitals and a neutron spin $\tau$ (=$\uparrow$ or $\downarrow$). 
In Eq.~(\ref{aobasis}), the index $m$($n$) is an abbreviation of the 
AO ($i,p_k,\tau$). The intrinsic basis functions 
with the full anti-symmetrization ${\cal A}$ are projected to the 
eigenstate of the total spin $J$, its intrinsic angular projection $K$ and 
the total parity $\pi$ by the projection operator $\hat{P}^{J^\pi}_K$. 

The total wave function is finally given by taking a 
superposition over $S$ and $K$ as 
\begin{eqnarray}
\Psi^{J^\pi}
\;&=&\;
\int dS\;\sum_{iK}\;C^K_i(S)\;\Phi^{J^\pi K}_i(S)
\label{adiwf}
\end{eqnarray}
with $i\equiv$ ($m$, $n$). The coefficients $C^K_i(S)$ are determined 
by solving a coupled channel GCM (Generator Coordinate Method) 
equation \cite{HORI}. If we fix the generator coordinate $S$ and 
diagonalize the Hamiltonian with respect to $i$ and $K$, we obtain 
the energy eigenvalues as a function of $S$, which we call the adiabatic 
energy surfaces (AES). 

In the present calculation, we used the Volkov No.2 and the 
G3RS for the central and the spin-orbit part of the 
nucleon-nucleon (NN) interaction, respectively. 
The parameters in the NN interactions are modified from those in 
Ref.~\cite{GTCM1} so as to reproduce the threshold of 
$\alpha$+$^6$He$_{g.s.}$ and the excitation energy of the $^6$He(2$_1^+$) 
state \cite{GTCM2}. This is because the reproduction of the threshold 
is essentially important in the treatment of the scattering problem.

The Majorana (M), Bartlett (B) and Heisenberg (H) 
exchanges in the central part are fixed to M=0.643, B=$-$H=0.125, while 
the strength of the spin-orbit force is chosen to 3000 MeV for the 
repulsive part and 2000 MeV for the attractive part. 
The radius parameter $b$ of HO wave functions for $\alpha$ clusters and 
valence neutrons is commonly taken as 1.46 fm. We included all the AO 
configurations of two neutrons that can be constructed by the $0p$-orbitals.

\begin{figure}
  \includegraphics[width=5cm]{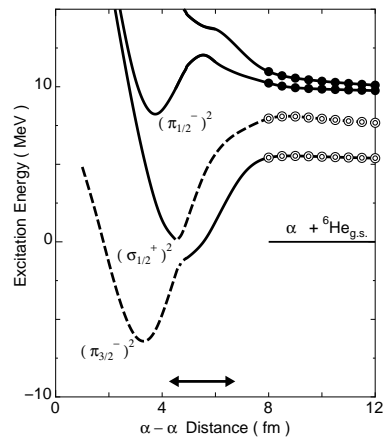} 
\vspace{-2mm}
\caption{
Adiabatic energy surfaces for $J^\pi$=0$^+$. 
The lowest and third local minima 
have a dominant configurations of ($\pi_{3\slash2}^-$)$^2$ and 
($\pi_{1\slash2}^-$)$^2$, respectively, while the 
second one has a configutaion of the distorted ($\sigma_{1\slash2}^+$)$^2$ 
configuration (See text for details).
The lowest and the second surfaces with the double circles
are the dinuclear states of [$\alpha$+$^6$He(0$_1^+$)]$_{L=0}$ and 
[$\alpha$+$^6$He(2$_1^+$)]$_{L=2}$, respectively, while 
those with the solid circles have the configurations of 
[$^5$He(3$\slash$2$^-$)+$^5$He(3$\slash$2$^-$)] with ($IL$)=(00) and 
(22), respectively. 
}
\label{0+sur}
\end{figure}

The AES with $J^\pi$=0$^+$ is shown in Fig.~\ref{0+sur}. 
There appear three local minima at the short distance region of 
the AES.  The adiabatic states (AS) at the lowest and 
third minima have the molecular orbital configurations of 
($\pi_{3\slash2}^-$)$^2$ and 
($\pi_{1\slash2}^-$)$^2$, respectively \cite{ITA,GTCM1,GTCM2}. 
The AS around the second minimum has a dominat configuration of 
($\sigma_{1\slash2}^+$)$^2$ \cite{ITA,4body2}, but the one particle excited 
configuration, ($\sigma_{1\slash2}^+\pi_{1\slash2}^+$), is also 
strongly mixed. The latter configuration has a spin triplet structure 
and hence, the coupling between them is induced by the two-body 
spin-orbit interaction. 

It is well known that the simple $\sigma_{1\slash2}^+$ orbital 
is not sufficient to describe both the 0$_2^+$ state in $^{10}$Be 
and the $1\slash2^+$ state in $^9$Be as discussed in Ref.~\cite{ITA}. 
Itagaki {\it et al.} shows that spin-orbit interaction 
generates the strong coupling betweeen the distorted 
($\sigma_{1\slash2}^+$)$^2$ configuration with the spin-triplet 
configuration and the pure ($\sigma_{1\slash2}^+$)$^2$ one, which 
plays very important role for lowering the 0$_2^+$ state in 
$^{10}$Be \cite{ITA}. Therefore, the present result is consistent 
to that discussed in Ref.~\cite{ITA} and it is 
reasonable to describe the intrinsic structure of $^{10}$Be(0$_2^+$). 

\begin{figure}
  \includegraphics[width=5cm]{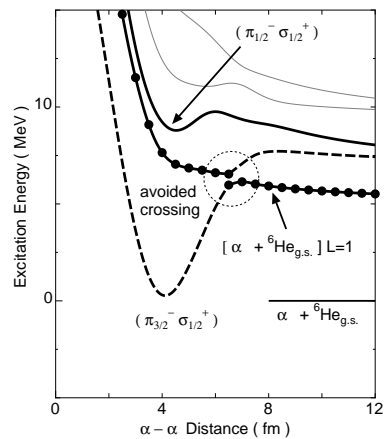}
\vspace{-2mm}
\caption{
The same as Fig.~\ref{0+sur} but for the negative parity 
states ($J^\pi$=1$^-$). The dashed (solid) surfaces 
has a dominant component of ($\pi_{3\slash2}^-\sigma_{1\slash2}^+$)$_{K=1}$ 
(($\pi_{1\slash2}^-\sigma_{1\slash2}^+$)$_{K=1}$) around the local 
minimum, while it is smoothly connected to the dinuclear channel of 
[$\alpha$+$^6$He(2$_1^+$)] with $L=1$ ($L=3$) at an asymptotic region. 
The surface with a solid circles has a dominant component of the 
[$\alpha$+$^6$He(0$_1^+$)]$_{L=1}$ channel. 
}
\label{1-sur}
\end{figure}

At the asymptotic region ($S\geq$ 8 fm) where two $\alpha$-cores 
are completely separated, the valence neutrons are localized around one 
of the $\alpha$ cores. The localization of the orbitals leads to the formation 
of the dinuclear channels such as [$^4$He+$^6$He($I$)]$_{L}$ 
(double circles) and [$^5$He($I_1$)+$^5$He($I_2$)]$_{IL}$ 
(solid circles), in which 
individual channels are specified by the intrinsic spin of the clusters 
($\mathbf{I_1}$, $\mathbf{I_2}$), the channel spin 
$I$ ($\mathbf{I}$=$\mathbf{I_1 + I_2}$) and the relative angular momentum 
between clusters, $L$. 
The asymptotic energy position of the lowest AES is higher 
by about 5 MeV than the $\alpha$+$^6$He$_{g.s.}$ threshold. This 
is because the relative motion between clusters in Eq.~(\ref{aobasis}) 
is described by the locally peaked Gaussian \cite{HORI} and hence, 
its kinetic energy contributes to the AES in the asymptotic region. 

The structural changes occur smoothly between the molecular orbital 
region and the dinuclear channels region in passing through the 
intermediate region shown by the arrow in Fig.~\ref{0+sur} 
\cite{GTCM1,GTCM2}. In the intermediate coupling region, we can see 
the level crossing between the surface of 
($\pi_{3\slash2}^-$)$^2$ (dashed curve) and the second 
one ($\sigma_{1\slash2}^+$)$^2$ (solid curve). The energy splitting 
at the crossing point is about 1.5 MeV. 


In Fig.~\ref{1-sur}, the AES for the $J^\pi$=1$^-$ state is shown. 
The configurations of the valence neutrons smoothly changes 
in the AES for the $\alpha-\alpha$ distance parameter 
except for the curves with the solid circles. 
The AS along this surface has an almost pure-component of the 
[$\alpha$+$^6$He(0$_1^+$)]$_{L=1}$ channel.  Thus, this AES is not molecular 
orbitals but the dinuclear state in the whole regions of $\alpha-\alpha$ 
distance. Because of the different character of the lowest two orbitals, 
the AES behaves different as a function of $\alpha-\alpha$ distance. 
In contrast to the results of the $J^\pi=0^+$ states shown 
in Fig.~\ref{0+sur}, this causes an clear avoided-crossing at $S$=6 fm as 
shown by the dotted circles in Fig.~\ref{1-sur}. 
The energy splitting at the crossing point 
is about 0.5 MeV which is smaller than that in $J^\pi$=0$^+$ ($\sim$1.5 MeV). 
This means that, in the negative parity state, the change of the intrinsic 
structure is much sharper in the distance than the case of 
the positive parity state.

\begin{figure}
  \includegraphics[width=5cm]{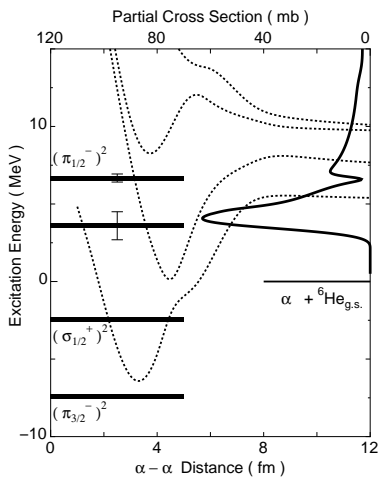}
\vspace{-2mm}
\caption{
Energy spectra for $J^\pi$=0$^+$. 
The error bar means the decay width of the resonance states. 
The solid curve shown in the right part is the partial cross section  
for the inelastic scattering [$\alpha$+$^6$He$_{g.s.}$]$_{L=0}$ 
$\rightarrow$ [$\alpha$+$^6$He(2$_1^+$)]$_{L=2}$ with a respective 
scale at the up-most axis. The AES in 
Fig.~\ref{0+sur} are shown by the dashed curves. 
}
\label{esp0+}
\end{figure}

\begin{figure}
  \includegraphics[width=5cm]{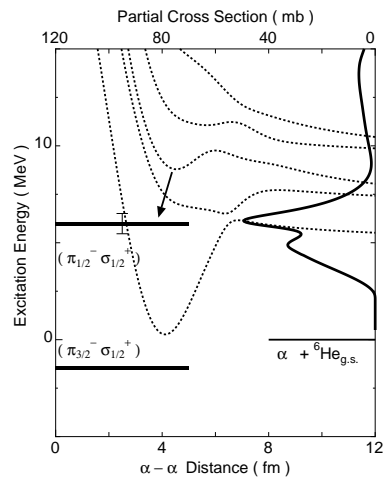}
\vspace{-2mm}
\caption{
The same as Fig.~\ref{esp0+} but for $J^\pi$=1$^-$. 
In the right part, the partial cross section 
for the inelastic scattering of [$\alpha$+$^6$He$_{g.s.}$]$_{L=0}$ 
$\rightarrow$ [$\alpha$+$^6$He(2$_1^+$)]$_{L=1}$ 
is shown by the solid curve. 
}
\label{esp1-}
\end{figure}

To take into account the excitation of the relative motions 
between two $\alpha$-cores, we solve the GCM equation by employing 
the AS from $S$=1 fm to $S$=70 fm with the mesh of 0.5 fm. 
The calculated energy spectra of $J^\pi$=0$^+$ and 
$J^\pi$=1$^-$ are shown in Figs.~\ref{esp0+} and \ref{esp1-}, respectively. 
In solving GCM, we apply the Absorbing-kernels in the 
Generator Coordinate Method (AGCM) in which the absorbing 
boundary condition is imposed outer region of the total system 
\cite{AGCM}. Due to the absorbing boundary, the resonance poles 
can be clearly identified in the complex energy plane. 

In the positive parity, the 0$_1^+$, 0$_2^+$ and 0$_4^+$ states 
are the poles corresponding to the respective local minima in the 
AES, having the molecular orbital structures. Therefore, we 
should call these states the ``adiabatic poles'', because they can 
be realized as the local minima in the AES. On the other hand, 
there is no local minimum corresponding to the 0$_3^+$ state and 
it is generated by the radial excitation on the distance parameter 
$S$. Thus, the 0$_3^+$ state should be called as the 
``radially-excited poles'', although it is the pole generated by a linear 
combination of the AS. The structure of 0$_3^+$ is very different 
from other lower adiabatic poles, because it is orthogonalized to the 
adiabatic poles and is spatially extended. 
The wave function in 0$_3^+$ has an enhanced component of 
[$\alpha$+$^6$He(2$_1^+$)]$_{L=2}$ at the surface region and 
hence, it is different from the molecular orbital configuration. 

In $J^\pi$=1$^-$, we have identified two adiabatic 
poles corresponding to the two local minima of 
($\pi^-_{j_z}\sigma^+_{1\slash2}$)$_{K=1}$. The lower pole 
($j_z=3\slash2$) is generated from the linear combination of the AS 
around the lowest local minimum, while the higher one ($j_z=1\slash2$) 
is originated from the minimum in the third AES as indicated by the 
arrow in Fig.~\ref{esp1-}.

In the present calculation, we identify no resonance 
corresponding to the $\alpha$+$^6$He$_{g.s.}$ AES, although 
its appearance is discussed within the bound state approximation 
in Ref.~\cite{GTCM1}. Therefore, the $\alpha$+$^6$He$_{g.s.}$ 
cluster configuration will not be stabilized as a resonance 
pole under the present condition reproducing the respective 
threshold energy. Since the $\alpha$+$^6$He$_{g.s.}$ 
AES has a flat shape over the wide range of the distance 
\cite{GTCM1}, the stablility of this configuration becomes 
quite sensitive to the energy position of the threshold. 



Let us consider the difference in the $J^\pi$=0$^+$ and 
1$^-$ in relation to the inversion doublet structure. 
The ($\pi_{3\slash2}^-$)$^2$ AES in $J^\pi$=0$^+$ and the 
$\alpha$+$^6$He$_{g.s.}$ AES in $J^\pi$=1$^-$ originally form the 
inversion doublet of $\alpha$+$^6$He$_{g.s.}$ with $K^\pi$=0$^\pm$. 
The $J^\pi$=0$^+$ partner couples to the symmetric $^5$He+$^5$He 
configuration and hence, it is strongly distorted to the molecular 
orbital as the distance $S$ gets closer. 
In the $J^\pi$=1$^-$ surface, however, the dinuclear 
configuration of $\alpha$+$^6$He$_{g.s.}$ is well developed, which 
is similar situation to the $^{20}$Ne=$\alpha$+$^{16}$O system 
\cite{PDBL}. Therefore, the appearance of the avoided crossing 
in the negative parity has a close connection to the formation of the 
inversion doublet in the compound system of $^{10}$Be. 

We next discuss how the AES profile and the level scheme appear in 
the low energy reactions. We show the partial cross section for 
the inelastic scattering to $^6$He(2$_1^+$) excitation.  
In solving the scattering problem, we employ the Kohn-Hulth\'en-Kato 
(KHK) variation method \cite{KHK} where the AO basis in Eq.~(\ref{aobasis})
are transformed to the asymptotic channel wave function. 
The present calculation is equivalent to the usual coupled-channels 
calculation including the channels of 
$\alpha$+$^6$He(0$_1^+$,0$_2^+$,2$_1^+$,2$_2^+$,1$^+$) and 
$^5$He(3$\slash$2$^-$,1$\slash$2$^-$)+$^5$He(3$\slash$2$^-$,1$\slash$2$^-$). 
Our calculation thus includes much more channel components than 
the previous studies of  $\alpha$+$^6$He cluster model \cite{DES}. 
In this calculation, the maximum $S$ is changed from 70 fm to 12 fm 
and the channel wave function is matched to the scattering Coulomb 
wave function at a matching radius $R_C$=11.7 fm.

The calculated partial inelastic cross-sections are shown in the 
right part of Figs.~\ref{esp0+} and \ref{esp1-}. In $J^\pi$=0$^+$, 
a strong peak appears at $E_{c.m.}$$\sim$ 3 MeV, although there is no 
definite avoided crossing in the AES. This is due to the effect of the 
radially-excited pole, 0$_3^+$, which include the large component of the 
exit 2$_1^+$ channel. We can also see the enhancements at $E_{c.m.}$
$\sim$ 7 MeV, which nicely coincident to the adiabatic pole of 
($\pi_{1\slash2}^-$)$^2$, 0$_4^+$, but it 
is much smaller than that generated by the radially-excited poles. 

\begin{figure}
  \includegraphics[width=4cm]{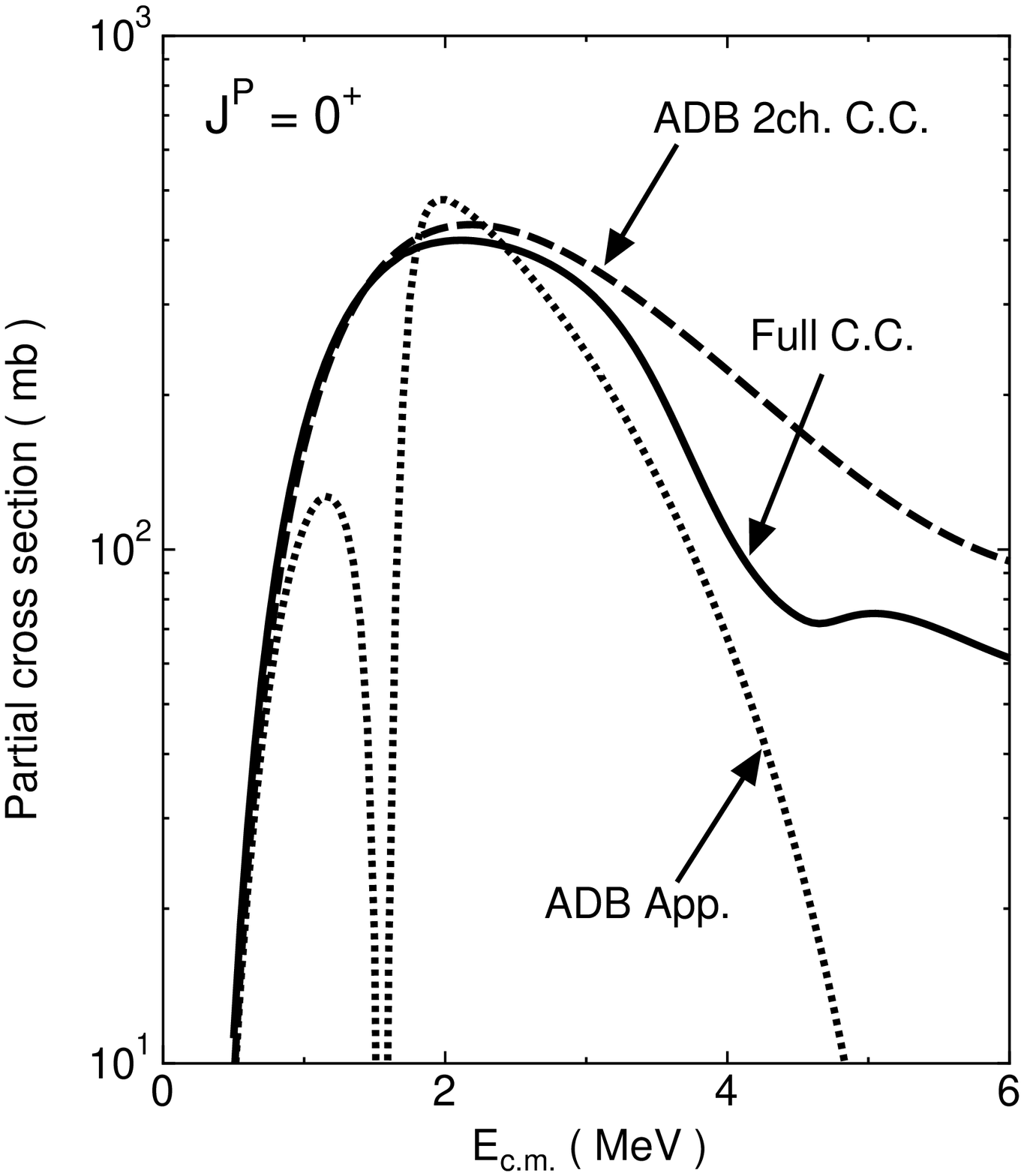}
  \includegraphics[width=4cm]{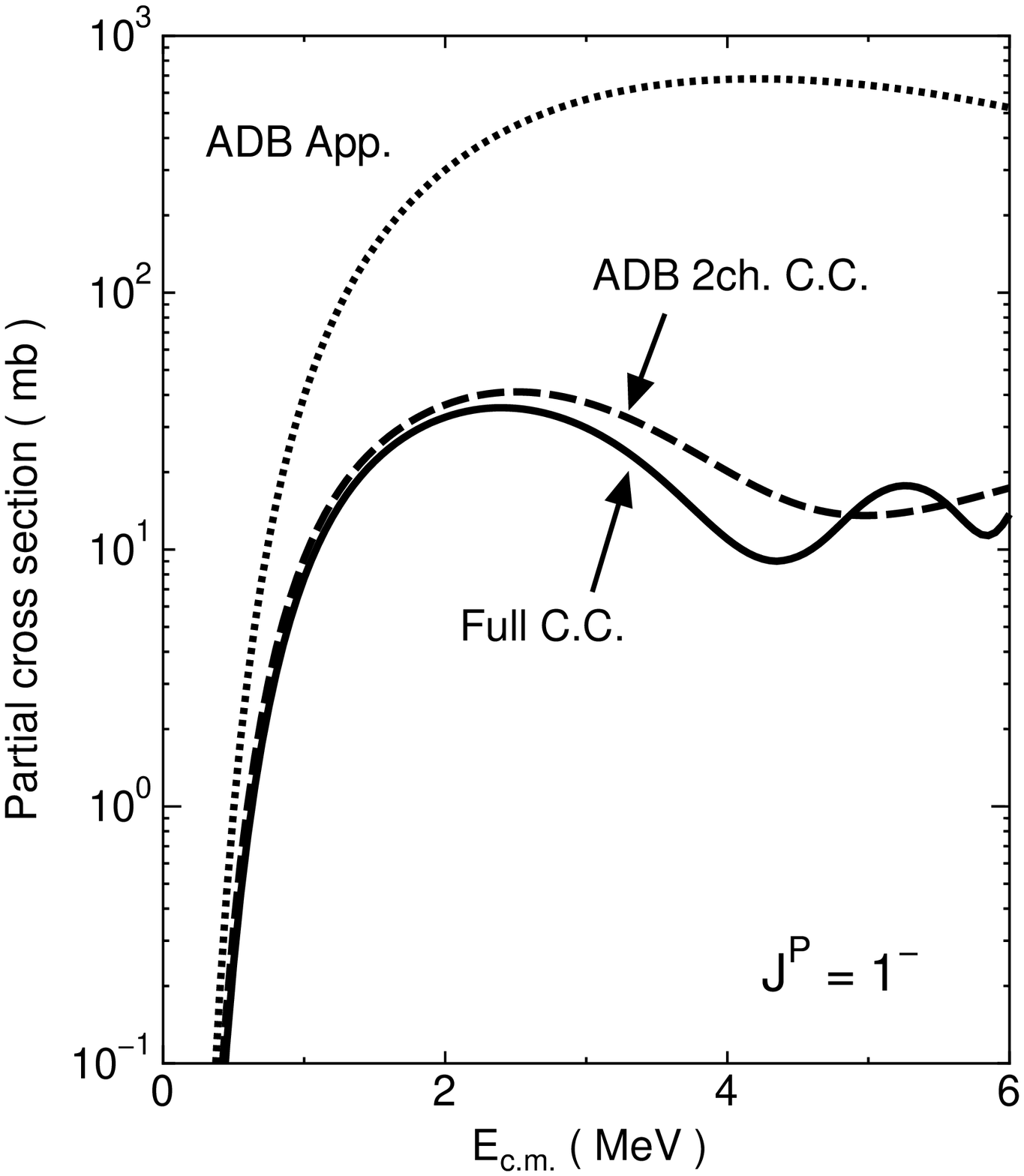}
\vspace{-2mm}
\caption{
The partial cross section of the $\alpha$+$^6$He$_{g.s.}$ elastic 
scattering. The left and right panels show the results for 
$J^\pi$=0$^+$ and $J^\pi$=1$^-$, respectively. 
The dotted curve shows the result of the adiabatic approximation, 
while the solid and dashed ones show that of two AS coupled-channel 
and that of the full coupled-channels, respectively, 
}
\label{el}
\end{figure}

In $J^\pi$=1$^-$, the strong enhancement can also be seen at 
$E_{c.m.}$$\sim$ 6 MeV, nevertheless there is no pole in the incident 
and exit channels. The energy of the enhancement is quite close to that of 
the avoided crossing at $S$=6 fm. To investigate the origin of this 
enhancement, we have solved the coupled-channel problem between the 
lowest two AS by employing the adiabatic Kohn-Hult\'en-Kato 
method that will be explained in the next paragraph. 
In such calculation, we have confirmed that this peak is generated by 
the coupling between the lowest two AS, nevertheless there appears 
no poles. Therefore, we can conclude that this inelastic 
peak is due to the L-Z transition at the avoided crossing. 
Though the ($\pi_{1\slash2}^-\sigma_{1\slash2}^+$) 
resonance is located close to the cross section peak, it weakly couples 
to the incident and exit channels. This pole is found to just generate 
the kink at a slightly lower energy than the peak position. 


To see the connection between the AS and the scattering 
process in a transparent way, we formulate the adiabatic KHK (AKHK) 
method in which individual AS are employed as the basis functions 
in solving the scattering problem. In the following, we briefly explain 
the formulation of the AKHK. First, we define the $a$-th AS at a 
distance $S$ by
\begin{eqnarray}
\Psi_{AS}^{J^\pi a}(S)\;&\equiv&\;
\sum_{i K}D^a_{i K}(S)\;\Phi^{J^\pi K}_i(S)\nonumber\\
\;&=&\;
\sum_{\beta}F^a_{\beta}(S)\;\Phi^{J^\pi \beta}_{CH}(S)\;\;,
\\
\label{adswf}
\Phi^{J^\pi \beta}_{CH}(S)
\;&=&\;{\cal A}\left\{
\left[\left[\varphi_{1I_1}(\xi_1)\otimes\varphi_{2I_2}(\xi_2)\right]_I
\otimes Y_L(\mathbf{\hat{R}})\right]_{J^\pi}\right.
\nonumber\\
\;&\times&\;\left.\chi_{L}(R,S)\right\}\;\;\;(\beta\equiv I_1I_2IL)\;\;,
\label{chwf}
\end{eqnarray}
where $\Phi^{J^\pi K}_i(S)$ is the AO basis given by 
Eq.~(\ref{aobasis}). In the second line, the AS is expanded 
in terms of the channel (CH) wave function, 
$\Phi^{J^\pi \beta}_{CH}(S)$. Eq.~(\ref{chwf}) shows 
the explicit expression of $\Phi^{J^\pi \beta}_{CH}(S)$ which is 
constructed from the angular momentum coupling among the 
internal states of $i$-th nucleus $\varphi_{iI_iM_i}(\xi_i)$ 
and the spherical harmonics $Y_{LM}(\mathbf{\hat{R}})$ with 
the relative coordinate $\mathbf{R}$. In the last line of 
Eq.~(\ref{chwf}), $\chi_L(R,S)$ denotes the locally peaked 
Gaussian with the peak position of $R\sim S$. 

The mixing coefficients in the $a$-th AS, $F^a_\beta(S)$, 
satisfies the following relation 
\begin{eqnarray}
\lim_{S\rightarrow\infty}F^a_\beta(S)\;\sim\;
\delta_{\beta,\alpha}\;\;.
\label{reladas}
\end{eqnarray}
Eq.~(\ref{reladas}) means that the $a$-th AS becomes 
a specific channel $\alpha$ at an asymptotic region 
($S\rightarrow\infty$), although 
a various channel components are strongly mixed in the internal 
region ($S\sim$ small). 
In solving the scattering problem 
with the basis of the AS, therefore, only a specific channel 
$\alpha$ satisfying Eq.~(\ref{reladas}) should be transformed into the 
scattering basis function, because the respective radial function 
$\chi_L(R,S)$ do not satisfy the scattering boundary condition. 

The transformation can be done by utilizing the 
KHK method, in which the localized basis 
functions are smoothly connected to the scattering Coulomb wave 
function at a matching radius $R_C$ \cite{KHK}. 
The $R_C$ should 
be taken to be sufficiently large value where all the channel components 
except for $\alpha$ are completely damped. 
The calculation of the matrix element for the AKHK basis can be easily done by 
the similar procedures shown in Ref.~\cite{KHK}. The details of 
the AKHK method will be shown in a forthcoming paper.  

In order to discuss the gross features of the non-adiabatic effects, 
we calculate the partial cross sections of the elastic scattering. 
In two panels of Fig.~\ref{el}, the results calculated from the AKHK 
method are shown. In this calculation, the $S_{max}$ and the $R_C$ 
are takne to be the same values in calculating the inelastic cross 
section. In both panels, the dotted curves 
show the pure adiabatic approximation. That is, only the AS along 
to the lowest AES is employed in solving the scattering problem. 
The solid curves show the solution of the full coupled channel (CC) 
in which all the AS are included. 

In the result of $J^\pi$=0$^+$, the adiabatic approximation 
quite nicely simulates the full CC solution in the low-energy region. 
Furthermore, the adiabatic approximation simulates the gross behavior 
of the full CC solution up to about 4 MeV except for the kink just below 
2 MeV. This means that the elastic scattering mainly proceeds along 
to the lowest AES. The dashed curve shows the result in which the lowest 
two AS are coupled. The coupling effect improves the adiabatic 
approximation. The dashed curve deviates from the solid one in the 
higher energy than 3 MeV, but the difference is not so large. 

The results of $J^\pi$=1$^-$ is drastically 
different from those of $J^\pi$=0$^+$. The validity of the 
adiabatic approximation is limited only in the region of the small 
cross section. Furthermore, the non-adiabatic coupling between the 
lowest two AS strongly reduces the cross section of the adiabatic 
approximation, which amounts to about one order reduction. Therefore, 
the adiabatic approximation is quite poor for describing the scattering 
process. The main contribution of the non-adiabatic coupling is from 
the first excited AS, because the dashed curve is similar to the 
solid one. This is due to the appearance of the avoided crossing 
which can be clearly seen in Fig.~\ref{1-sur}.

In summary, we investigated the adiabatic properties of the 
$^{10}$Be=$\alpha$+$\alpha$+$N$+$N$ structures as well as the 
$\alpha$+$^6$He$_{g.s.}$ low-energy reactions, especially the 
non-adiabatic dynamics including the Landau-Zener (L-Z) transition. 
We achieved such an unified study of the structures and the reactions 
in the generalized two-center cluster model (GTCM). 
The scattering problem in the GTCM is solved by applying the 
Kohn-Hulth\'en-Kato (KHK) variation method and the adiabatic KHK 
(AKHK) method proposed in the present study. We have found the 
non-adiabatic enhancements in the inelastic scattering cross-section 
to the $\alpha$+$^6$He(2$_1^+$) in both the $J^\pi$=0$^+$ and 
$J^\pi$=1$^-$ states. However, their origins are very different to 
each other. In the $J^\pi$=0$^+$, the enhancements is due to the 
appearance of the radially-excited pole in the $^{10}$Be system. This 
is realized by the radial excitation of the $\alpha$-core's 
relative motions. In contrast, the enhancement in $J^\pi$=1$^-$ 
is originated from the L-Z level-crossing between the 
lowest two adiabatic energy surfaces. 
Such a difference of the non-adiabatic dynamics also 
affects the elastic scattering process. In the positive parity, the 
adiabatic approximation is good for describing the collision process, 
while it becomes wrong approximation in the negative parity. 

Next important step is to compare our result with recent experiments 
\cite{RAAB,SHIM} by extending the present application to the 
higher partial waves. In a comparison with experiments, we should 
be careful to optimize the nucleon-nucleon interaction so as to 
reproduce the observed energy spectra in $^{10}$Be together with 
the threshold of $\alpha$+$^6$He and $^5$He+$^5$He. 
Such extended studies are under progress. 

The author would like to thank Prof.\  K.\ Yabana for his valuable 
comments and careful check of the manuscript. He also would like to 
thank to Prof.\ K.\ Kato and Prof.\ Y.\ Abe for their valuable discussions 
and encouragement. This work was performed as a part of the "Research 
Project for Study of Unstable Nuclei from Nuclear Cluster Aspects " 
at RIKEN.

\end{document}